\documentclass{iau}
\usepackage{graphicx}
\title[Mapping DIBs]{Putting the Diffuse Interstellar Bands\\ on the map --
literally!}
\author[Jacco Th. van Loon]{Jacco Th. van Loon}
\affiliation{Lennard-Jones Laboratories, Keele University, ST5 5BG, UK\\
email: {\tt j.t.van.loon@keele.ac.uk}}
\pubyear{2013}
\volume{297}
\pagerange{1-5}
\jname{The Diffuse Interstellar Bands}
\editors{Jan Cami \& Nick Cox, eds.}
\begin{document}
\maketitle
\begin{abstract}
In a quest to further our understanding of the diffuse interstellar medium
(ISM) as well as the unidentified carriers of the diffuse interstellar bands
(DIBs), we are mapping DIBs across the sky using hundreds of hot stars as
background torches -- globular clusters (in particular $\omega$\,Centauri),
nearby stars in and around the Local Bubble, and stars within the Magellanic
Clouds. I describe the results so far obtained and our current experiments.
\end{abstract}
\firstsection
\section{Rationale}

While the carriers of the DIBs are not known, their correlations with better
known constituents of the ISM suggest they are likely of molecular origin and
reside in diffuse clouds and cloud surfaces (see the review by Daniel Welty,
these proceedings). We can thus already start using DIBs to trace special ISM
conditions, such as those related to the atomic--molecular transition or cloud
irradiation effects, that otherwise may be rather elusive; hoping meanwhile to
expose clues that help identify their carriers.

As no DIB has been seen in emission, we rely on background continuum sources
to trace their absorption. Fields of stars, such as bright nearby stars, stars
in open and globular clusters and stars in nearby resolved galaxies provide a
means to construct sparse maps of the DIB absorption across a variety of ISM
volumes. Compared to correlation diagrams, these maps add a spatial dimension,
even rendering a 3-dimensional picture if stars at different known distances
are observed.

\section{Experimental characteristics}

I here describe several of our recent and current DIB mapping experiments,
summarised in Table\,\ref{tab1}. It all started with a spectroscopic survey of
the post-main sequence stars in the ``nearby'' ($\sim5$ kpc) most massive
Galactic globular cluster, $\omega$\,Centauri. This programme employed the
powerful 2dF multi-object spectrograph on the 4-m Anglo-Australian Telescope,
which allows to place up to 400 fibres on positions on the sky within a huge,
$2^\circ$-diameter field. The hot horizontal branch stars in that sample,
further aided by their low metal content ($\sim0.03$ Z$_\odot$) and high
velocity ($\sim230$ km s$^{-1}$) with respect to the intervening ISM, turned
out to be useful in mapping the Ca\,{\sc ii}\,K absorption. I here show
(below) that in fact those blue spectra also display some DIBs. To do this
more adequately, we designed a dedicated experiment with the enhanced 2dF
facility (AA$\Omega$), allowing more spectral coverage at higher spectral
resolution.

\begin{table}
\begin{center}
\caption{Overview of spectroscopic DIB mapping experiments.}
\label{tab1}
{\scriptsize
\begin{tabular}{lrlrrll}
\hline
{\bf Target area}  & $N_\star$ & {\bf Facility} &
{\bf $\lambda/\Delta\lambda$} & {\bf S/N} & {\bf Main DIBs} &
{\bf Publications} \\
\hline
$\omega$\,Centauri    & 231$\in$1528 & AAT/2dF        &  2,000 &  $>50$ &
4428       & \cite[van Loon \etal\ (2007)]{vanloon07}; here \\
                      &          452 & AAT/AA$\Omega$ &  8,000 &    100 &
5780, 5797 & \cite[van Loon \etal\ (2009)]{vanloon09}; here \\
Tarantula Nebula      &         800  & VLT/FLAMES     & 10,000 &    100 &
4428, 6614 & \cite[van Loon \etal\ (2013)]{vanloon13}       \\
Magellanic Clouds     &       $>600$ & AAT/AA$\Omega$ &  8,000 & $>200$ &
5780, 5797 & Bailey \etal\ (in prep.)                       \\
Local Bubble    South &          238 & NTT/EFOSC2     &  8,000 &  1,000 &
5780, 5797 & Bailey \etal\ (in prep.)                       \\
\hspace{16.2mm} North &          333 & INT/IDS        &  2,000 &  1,000 &
5780, 5797 & Farhang \etal\ (in prep.)                      \\
\hline
\end{tabular}
}
\end{center}
\end{table}

Likewise, we have just published DIB maps of the Tarantula Nebula, the famous
``mini-starburst'' region in the Large Magellanic Cloud (at a distance of 50
kpc, and with $\sim0.4$ Z$_\odot$ and a recession velocity of $\sim270$ km
s$^{-1}$). This too was a ``by-product'' of a stellar astrophysics programme
-- \cite[Evans \etal\ (2011)]{evans11}. It utilised the FLAMES spectrograph,
offering less multiplexity and coverage but a more sensitive 8-m Very Large
Telescope in Chile. We have since conducted a dedicated survey of both the LMC
and the Small Magellanic Cloud (60 kpc, $\sim0.2$ Z$_\odot$ and $\sim160$ km
s$^{-1}$) with AA$\Omega$; Keele University Ph.D.\ student Mandy Bailey has
now finished the measurements from these spectra, and a publication of the
results is expected to appear in 2014.

Inspired by the atomic-line 3-D mapping of the Local Bubble and its environs
by \cite[Lallement \etal\ (2003)]{lallement03} and \cite[Welsh \etal\
(2010)]{welsh10} we have conducted an all-sky survey of DIB absorption towards
552 different early-type stars, mostly within a few hundred pc from the Sun.
In contrast to the above efficient mapping experiments, this meant obtaining
one spectrum at a time. Because interstellar absorption in such nearby stars
is often inconspicuous, we aimed at a signal-to-noise ratio of 1000:1 or
better; this incurs high demands on the calibration measurements as well. The
Southern component of the survey was carried out at the 3.5-m New Technology
Telescope in Chile, by Mandy Bailey, whilst the Northern component was added
by Sharif University of Technology Ph.D.\ student Amin Farhang using Iranian
time on the 2.5-m Isaac Newton Telescope in Spain. Publications arising from
this programme are expected to appear in 2014.

\section{Extra-planar gas in front of $\omega$\,Centauri}

The sightline towards $\omega$\,Centauri is relatively simple, at $15^\circ$
Galactic latitude, with low reddening of $E(B-V)\sim0.1$ mag (\cite[McDonald
\etal\ 2009]{mcdonald09}). Yet several components are seen in atomic
absorption (\cite[Wood \& Bates 1994]{wood94}, and references therein): hot
gas within the Local Bubble, neutral gas beyond it and including the
Carina--Sagittarius spiral arm, and warm extra-planar gas at greater distances
$\sim0.5$--1 kpc above the Galactic Plane.

Judging from the marginally-resolved kinematics of the Ca\,{\sc ii}\,K line in
our spectra, and various correlations, we concluded that the 5780 \AA\ DIB
absorption towards $\omega$\,Centauri originates mostly in the extra-planar
gas. The high Ca\,{\sc ii}/Na\,{\sc i} and 5780 \AA/5797 \AA\ DIB ratios in
those sightlines indicate a $\sigma$-type environment, implying a strong
radiation field or perhaps structures embedded within a hotter gas. However,
some more neutral $\zeta$-type clouds were seen too, displaying stronger 5797
\AA\ DIB absorption; these are likely associated with the intervening spiral
arm. Intriguingly, the DIBs were found to trace small-scale structure down to
parsec scales (or below), both in the warm extra-planar gas and neutral gas of
the Disc.

\begin{figure}
\begin{center}
\includegraphics[width=100mm]{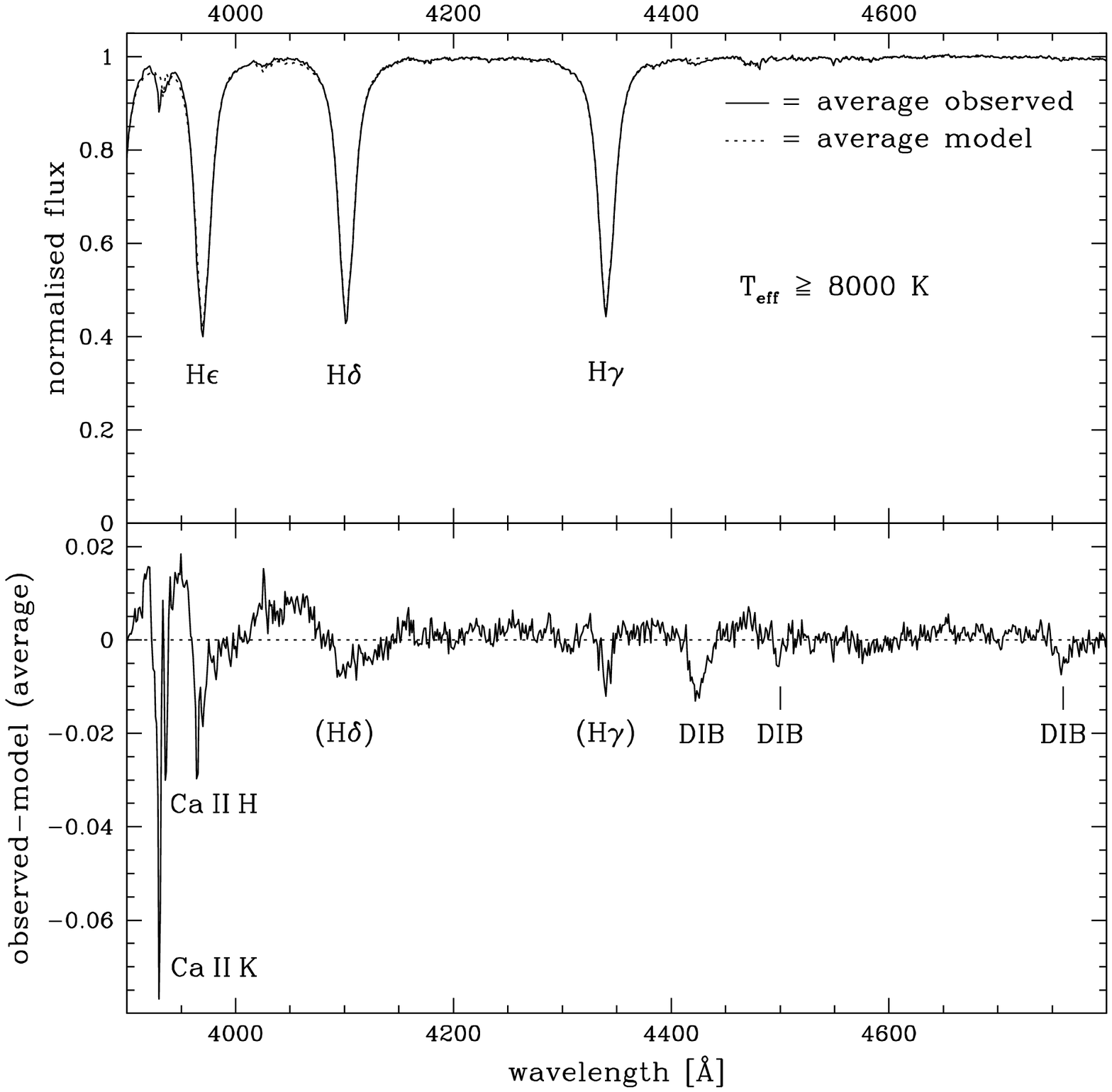}
\includegraphics[width=100mm]{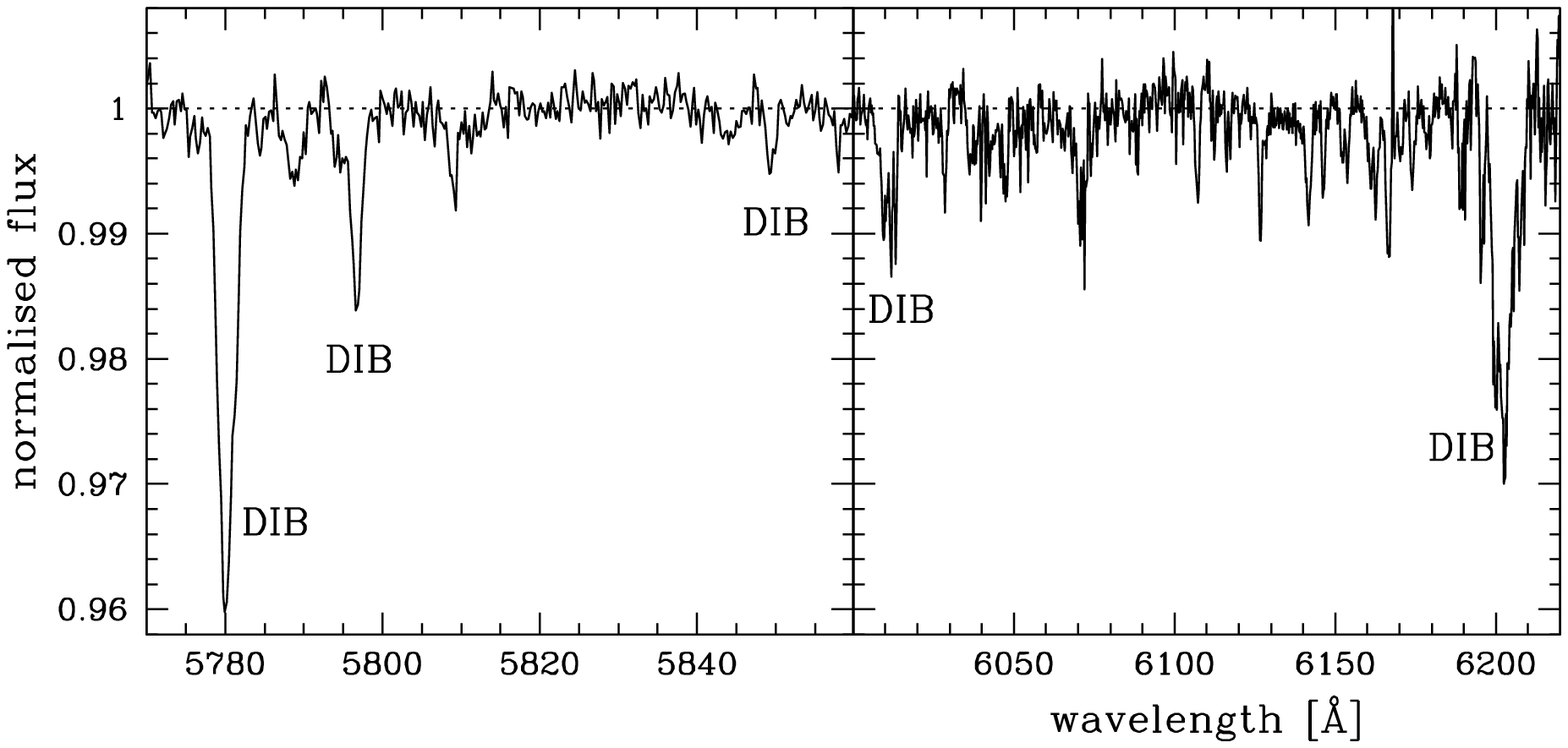}
\caption{DIBs detected in the spectra of hot stars in $\omega$\,Centauri. The
top panel is based on the survey of \cite[van Loon et al.\ (2007)]{vanloon07}
-- but we identify the DIBs for the first time here; the bottom panel is based
on \cite[van Loon et al.\ (2009)]{vanloon09} -- who only presented the 5780
and 5797 \AA\ DIBs.}
\label{fig1}
\end{center}
\end{figure}

A re-analysis of our spectra reveals some more DIBs (Figure\,\ref{fig1}): at
4428, 4500 and 4762 \AA\ in the hot-star spectra of \cite[van Loon \etal\
(2007)]{vanloon07}, and near 5850, 6010 and 6200 \AA\ in those of \cite[van
Loon \etal\ (2009)]{vanloon09}. With more targets available as well, there is
plenty of scope for more detailed studies in this interesting direction.

\section{The Magellanic Clouds and their Galactic foreground}

Our Magellanic DIB mapping exercise has thus far concentrated on the Tarantula
Nebula. This is a complex region characterised by multi-phase ISM including
molecular clouds (e.g., as traced in absorption by \cite[Tatton \etal\
2013]{tatton13}) and bubbles filled with hot gas, subject to a strong
radiation field from the many massive early-type stars in the central cluster
R\,136 and other young associations (cf.\ \cite[Evans \etal\ 2011]{evans11}).

We were able to separate the Magellanic absorption from that arising in the
Galactic foreground. The latter resembled the extra-planar environment seen in
the $\omega$\,Centauri maps. Small-scale structure within the LMC was probed
both spatially (down to $\sim10$ pc) and kinematically ($\sim10$ km s$^{-1}$;
up to six components in Na\,{\sc i}\,D). While the 5780 \AA/5797 \AA\ DIB
ratio in the Tarantula Nebula is generally high, regional variations occur in
the DIBs indicating particularly harsh environments towards the North and West
(in the direction of super-bubbles) and in the immediate vicinity of OB-type
associations, with more shielded conditions towards the South (where the
molecular ridge starts). The 4428 \AA\ DIB proved relatively strong in and
around the Tarantula Nebula, whilst the 5780 \AA\ DIB could be more strongly
confined to cloud skins.

Even if accounting for the reduced metal content of the LMC, the 5780, 5797
and 6614 \AA\ DIBs are weak. \cite[Welty \etal\ (2006)]{welty06} and \cite[Cox
\etal\ (2006]{cox06}\cite[, 2007)]{cox07} attributed this to the strong
radiation field. We proposed to also consider the nitrogen depletion in the
Magellanic Clouds (cf.\ \cite[van Loon \etal\
2010a]{vanloon10a},\cite[b]{vanloon10b}; see also Veronica Bierbaum's
contribution to these proceedings). I would also suggest to examine any
relation with the depletion of Poly-cyclic Aromatic Hydrocarbons (cf.\
\cite[Sandstrom \etal\ 2012]{sandstrom12}).

The relatively little studied set of DIBs at 4727, 4762 and 4780 \AA\ may
prove quite interesting, as we found some tentative evidence for differences
among these features between the Tarantula Nebula and the Galactic foreground.

While the results of our galaxy-wide Magellanic Cloud maps will be presented
elsewhere, we can already reveal that in the SMC very few sightlines show
DIBs. However, when it does it sometimes is fairly strong. We must consider
that, possibly, the conditions under which DIBs are seen might occur in
slightly different parts of clouds that have a reduced metal content (see also
the behaviour of interstellar ice -- \cite[Oliveira \etal\ 2013]{oliveira13}).

Our Magellanic Clouds surveys also offer a detailed view of the high Galactic
latitude DIB absorption, adding to our all-sky survey of nearby stars to probe
structure on scales of $\sim0.1$ pc in and around the Local Bubble.

\section{The Local Bubble}

\begin{figure}
\begin{center}
\includegraphics[width=135mm]{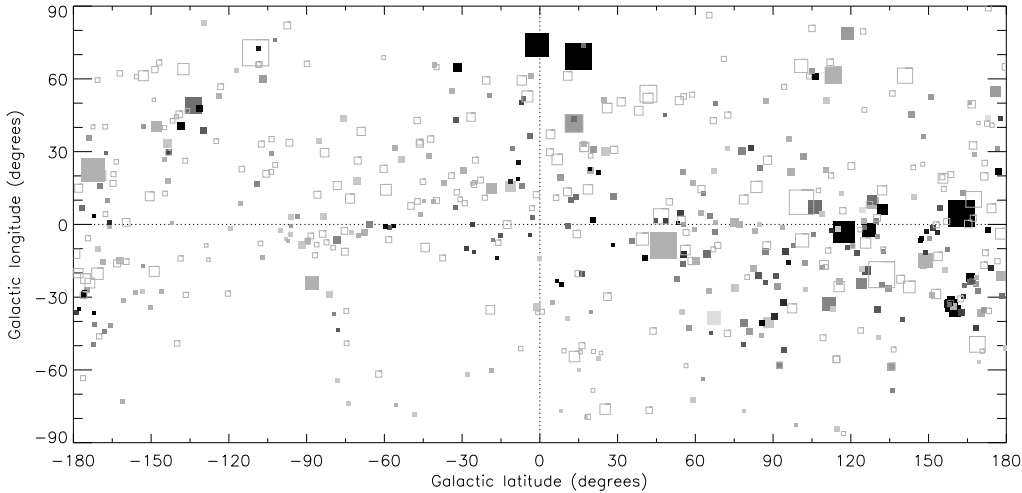}
\caption{Preliminary Galactic map of the Local Bubble measurements of the 5780
\AA\ DIB, combining the Southern (Bailey \etal) and Northern (Farhang \etal)
surveys. Darker symbols have stronger absorption, and larger symbols are
nearer; open symbols are non-detections.}
\label{fig2}
\end{center}
\end{figure}

We have managed to obtain reasonable all-sky coverage, maintaining also a
range in distance so we can probe the 3-D structure of the Local Bubble and
its surrounding neutral medium (Figure\,\ref{fig2}; Bailey \etal\ and Farhang
\etal, in preparation; see also Amin Farhang's contribution to these
proceedings).

While there is some concentration of 5780 \AA\ DIB absorption towards the
Galactic Plane for stars outside the Local Bubble (beyond $\sim100$--200 pc),
there also appear to be some DIB structures very nearby ($\ll 100$ pc), well
inside the Local Bubble and including some high-latitude clouds.

No less interesting are the weak or null detections, which are not restricted
to nearby or high-latitude stars. It is not uncommon for the 5780 \AA\ DIB to
be stronger in a nearby star seen in a direction close to that of a more
distant star with weaker DIB absorption. Indeed, there is a generally poor
correlation between DIB equivalent width and distance. The Milky Way as seen
in the 5780 \AA\ DIB appears rather patchy and porous.

\acknowledgments{I am grateful to the Universe for being so wonderfully
mysterious.}



\begin{thebibliography}{}
\bibitem[Cox \etal\ (2006)]{cox06}
{Cox, N.L.J., Cordiner, M.A., Cami, J., Foing, B.H., Sarre, P.J., Kaper, L.,
\& Ehrenfreund, P.} 2006, \textit{A\&A}, 447, 991
\bibitem[Cox \etal\ (2007)]{cox07}
{Cox, N.L.J., \etal} 2007, \textit{A\&A}, 470, 941
\bibitem[Evans \etal\ (2011)]{evans11}
{Evans, C.J., \etal} 2011, \textit{A\&A}, 530, A108
\bibitem[Lallement \etal\ (2003)]{lallement03}
{Lallement, R., Welsh, B.Y., Vergely, J.-L., Crifo, F., \& Sfeir, D.} 2003,
\textit{A\&A}, 411, 447
\bibitem[McDonald \etal\ (2009)]{mcdonald09}
{McDonald, I., van Loon, J.Th., Decin, L., Boyer, M.L., Dupree, A.K., Evans,
A., Gehrz, R.D., \& Woodward, C.E.} 2009, \textit{MNRAS}, 394, 831
\bibitem[Oliveira \etal\ (2013)]{oliveira13}
{Oliveira, J.M., \etal} 2013, \textit{MNRAS}, 428, 3001
\bibitem[Sandstrom \etal\ (2012)]{sandstrom12}
{Sandstrom, K.M., \etal} 2012, \textit{ApJ}, 744, 20
\bibitem[Tatton \etal\ (2013)]{tatton13}
{Tatton, B.L., \etal} 2013, \textit{A\&A}, 554, A33
\bibitem[van Loon \etal\ (2007)]{vanloon07}
{van Loon, J.Th., van Leeuwen, F., Smalley, B.S., Smith, A.W., Lyons, N.A.,
McDonald, I., \& Boyer, M.L.} 2007, \textit{MNRAS}, 382, 1353
\bibitem[van Loon \etal\ (2009)]{vanloon09}
{van Loon, J.Th., Smith, K.T., McDonald, I., Sarre, P.J., Fossey, S.J., \&
Sharp, R.G.} 2009, \textit{MNRAS}, 399, 195
\bibitem[van Loon \etal\ (2010a)]{vanloon10a}
{van Loon, J.Th., \etal} 2010a, \textit{AJ}, 139, 68
\bibitem[van Loon \etal\ (2010b)]{vanloon10b}
{van Loon, J.Th., Oliveira, J.M., Gordon, K.D., Sloan, G.C., \& Engelbracht,
C.W.} 2010b, \textit{AJ}, 139, 1553
\bibitem[van Loon \etal\ (2013)]{vanloon13}
{van Loon, J.Th., \etal} 2013, \textit{A\&A}, 550, A108
\bibitem[Welsh \etal\ (2010)]{welsh10}
{Welsh, B.Y., Lallement, R., Vergely, J.-L., \& Raimond, S.} 2010,
\textit{A\&A}, 510, A54
\bibitem[Welty \etal\ (2006)]{welty06}
{Welty, D.E., Federman, S.R., Gredel, R., Thorburn, J.A., \& Lambert, D.L.}
2006, \textit{ApJS}, 165, 138
\bibitem[Wood \& Bates (1994)]{wood94}
{Wood, K.D., \& Bates, B.} 1994, \textit{MNRAS}, 267, 660
\end{thebibliography}
\end{document}